\documentclass[11pt,a4paper]{article}
\usepackage{moriond_kellerbauer,graphicx}
\usepackage[numbers]{natbib}
\begin{document}
\vspace*{4cm}
\title{PRODUCTION OF COLD ANTIHYDROGEN WITH ATHENA\\
FOR FUNDAMENTAL STUDIES}

\author{\MakeUppercase{A.~Kellerbauer$^{1}$, %
M.~Amoretti$^{2,3}$, %
C.~Amsler$^{4}$, %
G.~Bonomi$^{1}$, %
P.~D.~Bowe$^{5}$, %
C.~Canali$^{2,3}$, %
C.~Carraro$^{2,3}$, %
C.~L.~Cesar$^{6}$, %
M.~Charlton$^{5}$, %
M.~Doser$^{1}$, %
A.~Fontana$^{7,8}$, %
M.~C.~Fujiwara$^{9}$, %
R.~Funakoshi$^{10}$, %
P.~Genova$^{7,8}$, %
J.~S.~Hangst$^{11}$, %
R.~S.~Hayano$^{10}$, %
I.~Johnson$^{4}$, %
L.~V.~J{\o}rgensen$^{5}$, %
V.~Lagomarsino$^{2,3}$, %
R.~Landua$^{1}$, %
E.~Lodi~Rizzini$^{12,8}$, %
M.~Macr\'{i}$^{2,3}$, %
N.~Madsen$^{11}$, %
G.~Manuzio$^{2,3}$, %
D.~Mitchard$^{5}$, %
P.~Montagna$^{7,8}$, %
H.~Pruys$^{4}$, %
C.~Regenfus$^{4}$, %
A.~Rotondi$^{7,8}$, %
G.~Testera$^{2,3}$, %
A.~Variola$^{2,3}$, %
L.~Venturelli$^{12,8}$, %
D.~P.~van~der~Werf$^{5}$, %
Y.~Yamazaki$^{9}$, %
N.~Zurlo$^{12,8}$}\\[0.5\baselineskip]}

\address{$^{1}$Department of Physics, CERN, 1211 Gen\`{e}ve 23, Switzerland \\
$^{2}$Dipartimento di Fisica, Universit\`{a} di Genova, 16146 Genova,
Italy\\
$^{3}$INFN Sezione di Genova, 16146 Genova, Italy\\
$^{4}$Physik-Institut, University of Zurich, 8057 Z\"{u}rich,
Switzerland\\
$^{5}$Department of Physics, University of Wales Swansea, Swansea
SA2~8PP, UK\\
$^{6}$Instituto di Fisica, Universidade Federal do Rio de Janeiro, Rio
de Janeiro 21945-970, Brazil\\
$^{7}$Dipartimento di Fisica Nucleare e Teorica, Universit\`{a} di
Pavia, 27100 Pavia, Italy\\
$^{8}$INFN Sezione di Pavia, 27100 Pavia, Italy\\
$^{9}$Atomic Physics Laboratory, RIKEN, Saitama 351-0198, Japan\\
$^{10}$Department of Physics, University of Tokyo, Tokyo 113-0033,
Japan\\
$^{11}$Department of Physics and Astronomy, University of Aarhus,
8000~Aarhus~C, Denmark\\
$^{12}$Dip. di Chimica e Fisica per l'Ingegneria e per i Materiali,
Universit\`{a} di Brescia, 25123 Brescia, Italy\\[0.5\baselineskip]
\textnormal{(ATHENA Collaboration)}\\[0.5\baselineskip]}

\maketitle%
\abstracts{Since the beginning of operations of the CERN Antiproton
Decelerator in July 2000, the successful deceleration, storage and
manipulation of antiprotons has led to remarkable progress in the
production of antimatter. The ATHENA Collaboration were the first to
create and detect cold antihydrogen in 2002, and we can today produce
large enough amounts of antiatoms to study their properties as well as
the parameters that govern their production rate.}

\newpage
\section{Introduction}

\begin{figure}
  \centering
  \includegraphics{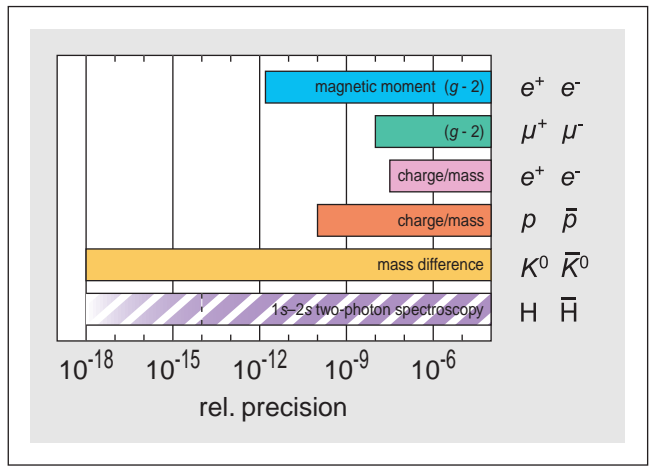}%
  \caption{Comparison of experimental CPT tests for various classes of
particle--antiparticle pairs with the \textit{potential} for a CPT test
with antihydrogen, based on the currently achieved precision in the
determination of the hydrogen 1$s$--2$s$ transition
frequency ($10^{-14}$)~\cite{bib:nier2000} and the limit due to its
natural linewidth ($10^{-18}$).%
\label{fig:CPT_comparison}}
\end{figure}

In 1957, the invariance of all laws of physics under the parity
transformation (the reversal of the spatial configuration) was shown to
be strongly violated in the weak decay~\cite{bib:wu__1957}. The failure
of such a fundamental symmetry, which was until then universally
believed to hold, reminded scientists that human intuition can be
treacherous in our quest to more fully describe nature. This
realization, along with the discovery of the more subtle violation of
CP symmetry (simultaneous reversal of charge and
space)~\cite{bib:chri1964}, has made physicists wary that the last of
the fundamental symmetries, CPT, may also eventually be shown to be
broken by sufficiently precise experiments. According to the CPT
theorem, which also establishes an intimate link between CPT symmetry
and Lorentz invariance, the simultaneous reversal of charge, space, and
time transforms any particle into its antiparticle~\cite{bib:paul1957}.
Particles and their antiparticles hence share many of their fundamental
properties (such as mass and half-life), and some other properties are
equal in magnitude and reversed in sign (such as charge and magnetic
moment). Any difference of those properties between
particle-antiparticle pairs would therefore be a clear proof of CPT
violation and an indication of physics beyond the Standard Model.
Spurred by the failure of C and CP and by recent theories that suggest
the possibility of CPT violation~\cite{bib:elli1992,bib:bluh1998},
experimental comparisons of particle-antiparticle pairs have been
performed on all classes of particles and to quite high precision. No
deviations have been observed to date. Figure~\ref{fig:CPT_comparison}
shows a comparison of some of those measurements (note the logarithmic
scale).
The most precise CPT test for leptons stems from the comparison of the
$g$~factors of the electron and the positron to $2 \times
10^{-9}$~\cite{bib:vand1987}. The best verification for baryons and
antibaryons is from the proton-antiproton charge-to-mass ratio measured
in a Penning trap~\cite{bib:gabr1999}. Finally, the mass difference of
the neutral $K$~meson and its antiparticle, with a relative uncertainty
of only $3 \times 10^{-19}$, yields the best verification of CPT
invariance for mesons (though this result is
model-dependent)~\cite{bib:hsiu2000}.

In the last two decades, immense progress has been made in the field of
high-precision atomic spectroscopy of the hydrogen atom. The transition
between the atomic ground (1$s$) and first excited (2$s$) state of
hydrogen can be measured to 2 parts in $10^{14}$~\cite{bib:nier2000}.
For this purpose, a cold (7~K) hydrogen beam is brought into overlap in
a resonant cavity with a laser beam whose wavelength is precisely
double that of the 1$s$--2$s$ transition, thereby allowing
(first-order) Doppler-free two-photon excitation. A quenching microwave
field, which is turned on alternatingly with the exciting laser, resets
the atoms from the metastable ($\tau = 122$~ms) 2$s$ state to the
short-lived ($\tau = 1.6$~ns) 2$p$ state whose decay to the ground
state by emission of a 121-nm photon is detected with a photomultiplier
tube. This achievement paves the way for a CPT test with hydrogen and
antihydrogen that could considerably improve the experimental limits
for both leptons and baryons. It must be pointed out, however, that the
actual quantity measured in such a comparison is the ratio of the
Rydberg constant to the anti-Rydberg constant. Since that ratio is a
function of the charges and the masses of the proton and the electron
as well as their antiparticles, possible CPT violating terms could
cancel out.

The second domain of fundamental tests that could be carried out with
antihydrogen concerns antimatter gravity. The Equivalence Principle of
General Relativity, in particular the universality of free fall, has
been well tested with ordinary matter~\cite{bib:adel2003}. Examples for
instruments used in these tests are torsion
balances~\cite{bib:heck2000} and atomic fountains~\cite{bib:pete1999}.
On the other hand, no macroscopic bodies of antimatter are available,
and precision tests with charged elementary anti-particles are hampered
by the relative strength of the electromagnetic force with respect to
gravity. This difficulty could be overcome with neutral antimatter
atoms, under the condition that confinement and cooling of anti-atoms
in a neutral-particle trap can be achieved. Possible deviations from
the Weak Equivalence Principle could arise from gravitational scalar or
vector fields. These effects could very well be stronger than possible
CPT violations, because the gravitational pull by ordinary matter on
antimatter does not constitute the CPT-symmetric situation to
matter--matter attraction.

Any high-precision measurement of a ground state property requires a
cold sample and long observation times. A large supply of the studied
species is helpful and in some cases, such as for experiments using a
beam of particles, absolutely required. It was therefore the declared
goal of the ATHENA Collaboration (and also of its direct competitor
ATRAP) to attempt to demonstrate the production of antihydrogen from
cold antiprotons and positrons trapped in a charged-particle trap and
to furthermore optimize the number of produced antihydrogen atoms and
study their properties. These constitute the main objectives of the
ATHENA experiment for its first phase, which started with the
commissioning of the CERN Antiproton Decelerator in 1999 and which will
end with its temporary shutdown at the end of 2004.

\section{Experimental Setup}

\begin{figure}
  \centering
  \includegraphics{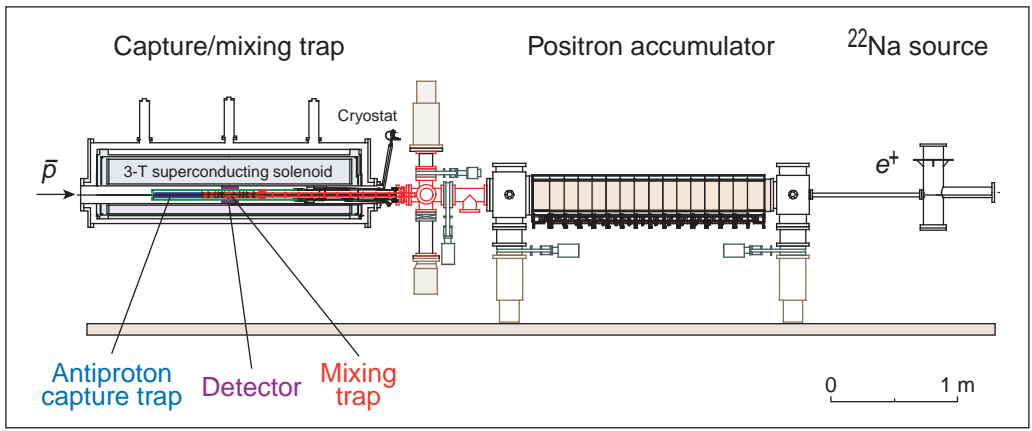}%
  \caption{Overview of the ATHENA apparatus. Shown on the left is the
superconducting 3-T solenoid magnet which houses the capture trap, the
mixing trap, and the antihydrogen annihilation detector. On the right,
the radioactive sodium source for the positron production and
the 0.14-T positron accumulation Penning trap.%
\label{fig:ATHENA_setup}}
\end{figure}

The layout of the ATHENA apparatus is based on an independent
preparation of cold antiprotons and positrons and their subsequent
mixing in a common trap region~\cite{bib:amor2004a}. The main element
of the experiment is a superconducting solenoid magnet which contains
the capture and mixing traps, two parts of a long cylindrical Penning
trap (diameter 25~mm, total length $\approx$1~m; see
Fig.~\ref{fig:ATHENA_setup}).
The magnet is operated at a field magnitude of 3~T. A cryostat with a
liquid-helium circuit independent from that of the magnet cools the
trap to a temperature of about 15~K.

The antiprotons are produced in the target area of the Antiproton
Decelerator (AD)~\cite{bib:heme1999}, where a thick iridium target is
bombarded with a 24-GeV/$c$ pulse of about $10^{13}$ protons. The
antiprotons are separated from the secondary products such as pions and
muons and 1--3$\times 10^{7}$ $\bar{p}$ are injected into the AD at a
momentum of $\approx$3.5~GeV/$c$. This antiproton bunch is then
decelerated in several stages and radially cooled by stochastic
(velocity $v \approx c$) and electron cooling ($v
\approx$~0.3--0.1$c$). After about 100~s, the pulse is extracted at
100~MeV/$c$ ($\approx$5~MeV kin.\ energy) and delivered to one of the
experiments. ATHENA performs the last deceleration step from the AD
final energy of about 5~MeV by letting the antiprotons pass through a
thin ($\approx$50~$\mu$m) degrader foil. The foil thickness is chosen
so as to maximize the fraction of $\bar{p}$ with an energy of less
than~5~keV, thereby allowing an efficient stopping by electrostatic
potentials. In the capture trap, the $\approx$$10^{4}$ antiprotons per
AD spill lose energy by collisions with a plasma of about $10^{8}$
electrons which have been loaded into the trap and allowed to cool.

In parallel, positrons are produced from a strong source of radioactive
$^{22}$Na, which decays to $^{22}$Ne by emitting highly energetic
$\beta$~radiation. The positrons are moderated in a layer of solid neon
and cooled and accumulated in the axial potential minimum of a
buffer-gas-filled 0.14-T Penning trap. After each accumulation cycle of
about 200~s, between $5 \times 10^{7}$ and $1.5 \times 10^{8}$
positrons are available for transfer to the recombination region.
There, the number, density, aspect ratio, and relative temperature of
the positron plasma can be non-destructively measured with a detection
system that resonantly excites and detects the first and second axial
plasma modes~\cite{bib:amor2003}.

\begin{figure}
  \centering
  \includegraphics{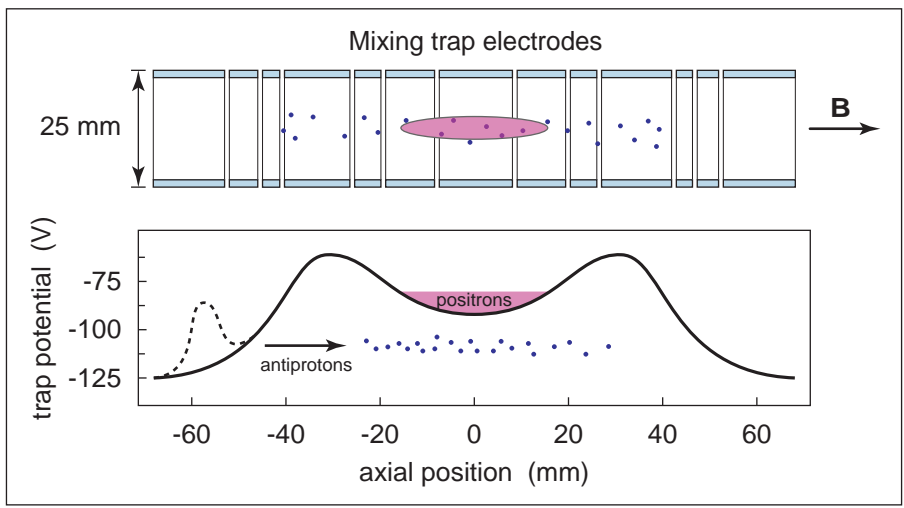}%
  \caption{Detailed sketch of the mixing trap, which is operated
in a nested-trap configuration~\cite{bib:gabr1988}. The bottom graph
shows the axial trap potential before (dashed line) and after (solid
line) the antiproton injection.%
\label{fig:Reco_trap}}
\end{figure}

The mixing of antiprotons and positrons takes place in a second
cylindrical Penning trap in the superconducting solenoid. This
so-called mixing trap is shown schematically in
Fig.~\ref{fig:Reco_trap}.
In order to simultaneously trap particles of opposite charge, the axial
electrostatic potential in the recombination region is operated in a
so-called nested-trap configuration~\cite{bib:gabr1988}. The central
well is first filled with the positrons, which cool to the ambient
temperature of 15~K with a time constant of about 0.5~s by synchrotron
radiation due to their high cyclotron frequency in the strong magnetic
field. The $\approx$$10^{4}$ antiprotons are initially transferred to a
small lateral well (dashed line in Fig.~\ref{fig:Reco_trap}) and then
launched into the nested-well region with a kinetic energy of about
30~eV. The antiprotons cool in collisions with the positrons and after
about 20--30~ms $\overline{\mbox{H}}$ production sets in. Neutral
antihydrogen atoms are no longer confined in the charged-particle trap
and leave the interaction region with a momentum essentially equal to
that of the antiproton just before recombination. They hit the trap
electrodes, where their constituents annihilate with their
ordinary-matter counterparts in quick succession. The mixing cycle
currently lasts about~70~s and is repeated every 250~s.

\begin{figure}
  \centering
  \includegraphics{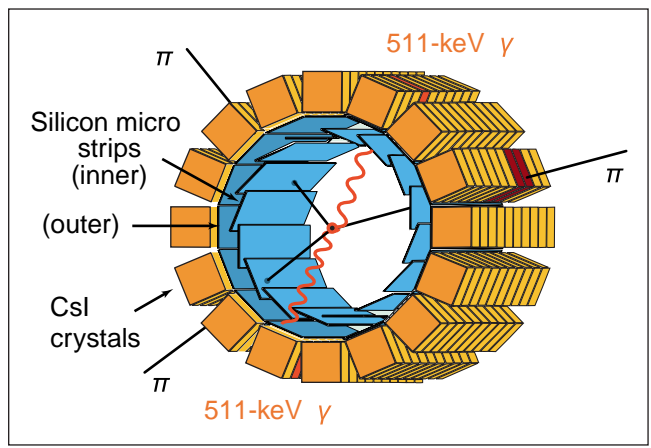}%
  \caption{Sketch of the ATHENA antihydrogen annihilation detector. With
its highly granular silicon strip and CsI crystal modules, it allows a
direct and unambiguous detection of $\overline{\mbox{H}}$ production.%
\label{fig:Detector}}
\end{figure}

The signal of the $\overline{\mbox{H}}$ annihilations is recorded with
the ATHENA antihydrogen detector, positioned coaxially with the mixing
trap within the magnet bore. It is a crucial component of this
experiment and sets it apart from its competitor. The active components
of the ATHENA detector are shown schematically in
Fig.~\ref{fig:Detector}.
It consists of a highly granular array of 8192 silicon strips in two
layers and 192 CsI crystals read out by avalanche photo diodes, all
within a radial extent of only 140~mm. The Si detectors record the
signal of the charged pions produced in the $\bar{p}$--$p$
annihilation; three or more hits in the outer Si layer trigger a
readout of the complete detector. Depending on the multiplicity of the
charged decay products, the Si detector hits can in most cases be used
to reconstruct the vertex of the annihilation. The resolution of this
reconstruction, about 4~mm, is limited by the fact that the curvature
of the charged-particle tracks cannot be reconstructed from only two
detector hits per track. The array of crystals detects the photons from
the $e^{+}$--$e^{-}$ annihilation, most of which ($> 95\%$) have a
multiplicity of two and are hence emitted back-to-back, \textit{i.e.}
with an opening angle of 180$^{\circ}$ as seen from the reconstructed
vertices from the charged particles.

\begin{figure}
  \centering
  \includegraphics{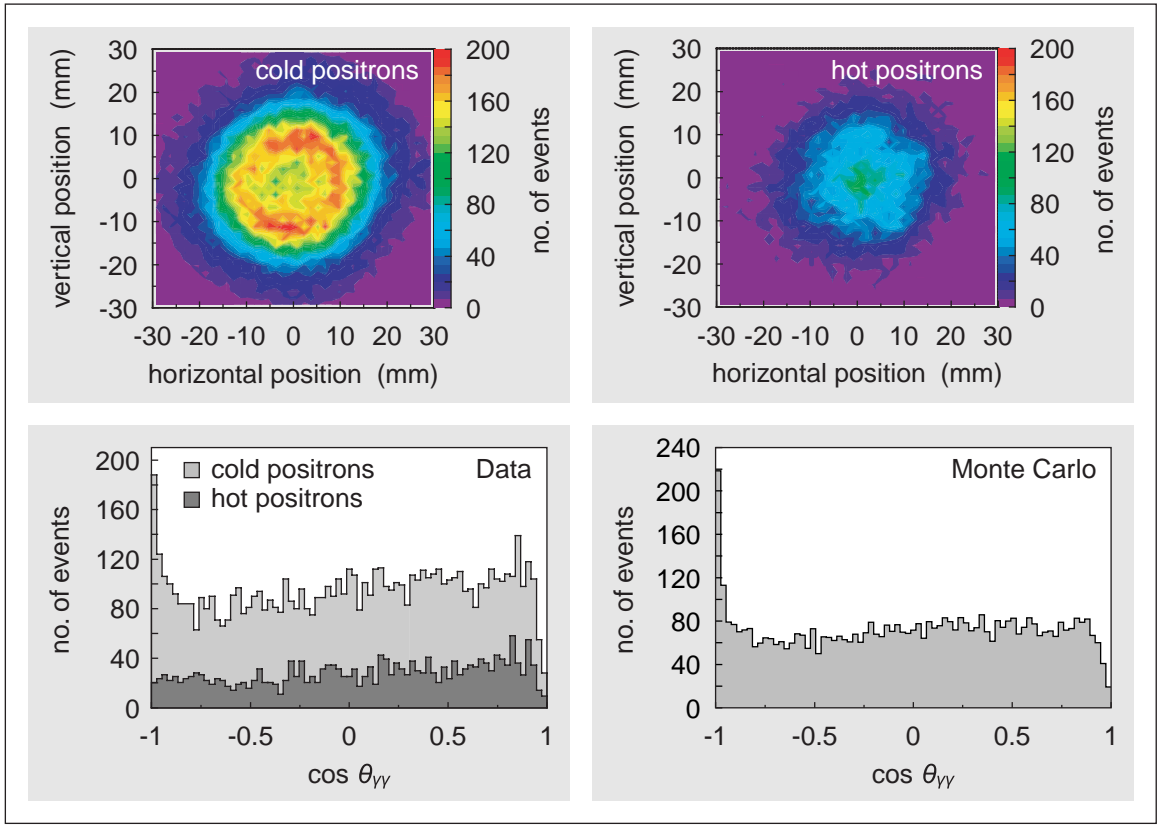}%
  \caption{Signal for the first production of cold antihydrogen with
ATHENA~\cite{bib:amor2002}. (top) Charged-pion vertex distribution as a
function of the azimuthal coordinates; (bottom) Opening-angle
distribution of the
photons recorded in coincidence with the charged-particle hits.%
\label{fig:Detector_signal}}
\end{figure}

Figure~\ref{fig:Detector_signal} shows the signal obtained with this
detector in a number of mixing cycles, as published in ATHENA's report
on the first production of cold antihydrogen in September
2002~\cite{bib:amor2002}.
The top panels show the number of recorded events from the silicon
detector as a function of the azimuthal coordinates. The highest number
of events, shown in red, is recorded at the trap electrodes at $r =
12.5$~mm as expected for antihydrogen annihilation. When the positron
plasma is heated to several 1000~K by an RF excitation of its axial
plasma modes, the $\overline{\mbox{H}}$ production is suppressed and
only a much lower number of events, from antiproton annihilations with
trapped residual gas ions or neutral contaminants, is detected (top
right panel). The lower panels of Fig.~\ref{fig:Detector_signal} show
the number of 511-keV $\gamma$ events recorded in the crystal detector
as a function of the opening angle $\theta_{\gamma\gamma}$. For cold
positrons, the data shows a clear peak for an opening angle of
180$^{\circ}$ which disappears when the positrons are heated. The right
panel illustrates the excellent agreement with a Monte Carlo
simulation. In Ref.~\citealp{bib:amor2002}, it was deduced from Monte
Carlo simulations that the 131 fully reconstructed events that
constitute the peak in the lower left panel of
Fig.~\ref{fig:Detector_signal} corresponded to a total number of
about~50\hspace{0.25em}000 produced $\overline{\mbox{H}}$~atoms.

In the months after ATHENA's initial report, the ATRAP Collaboration
also reported the production of cold antihydrogen~\cite{bib:gabr2002}.
Due to the lack of a position-sensitive $\overline{\mbox{H}}$
annihilation detector, ATRAP adopted an indirect detection scheme in
which antihydrogen that is emitted along the trap axis is re-ionized in
a strong electric field and the resulting antiprotons are captured in
an axial potential well. This allows a practically background-free
observation at the expense of a very small solid angle for detection.

A more complete analysis of our data from 2002, together with more
detailed Monte Carlo simulations, showed that the instantaneous trigger
rate from the silicon detector is a good proxy for antihydrogen
production, with 65\% of all triggers over the entire mixing cycle due
to annihilating antihydrogen atoms~\cite{bib:amor2004b}. A total of
about $2 \times 10^{6}$ antihydrogen atoms were produced by ATHENA in
2002 and 2003 combined.

\section{Recent Studies}

\subsection{Antiproton cooling}

\begin{figure}
  \centering
  \includegraphics{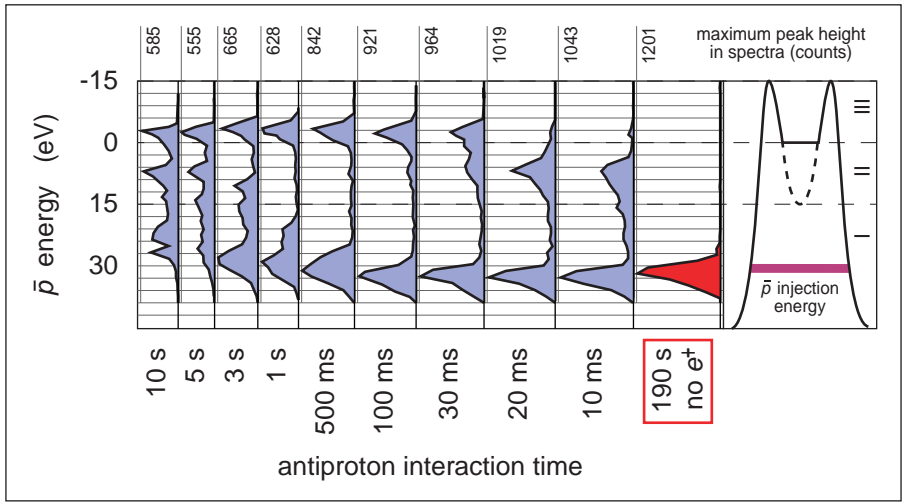}%
  \caption{Energy distribution of antiprotons, dumped from the nested
trap at varying times after injection~\cite{bib:amor2004c}.%
\label{fig:Antiproton_cooling}}
\end{figure}

As mentioned previously, the positron plasma confined in the 3-T
Penning trap cools rapidly ($\tau \approx 0.5$~s) to the temperature of
the surrounding electrodes by synchrotron radiation. The cooling of the
antiprotons in this cold $e^{+}$ plasma is a much more complicated
process because it is caused by Coulomb interactions with the positron
plasma and because of the unknown radial extent of the $\bar{p}$ cloud.
We have conducted a series of experiments to study this cooling
process, in which the nested wells of the mixing trap were slowly
emptied at varying times after the $\bar{p}$ injection, allowing a
measurement of the $\bar{p}$ energy distribution at that
time~\cite{bib:amor2004c}. The result of these measurements is shown in
Fig.~\ref{fig:Antiproton_cooling} (only the dump of the left lateral
antiproton well is shown).
The figure shows the presence of two distinct cooling processes with
very different time constants. A large portion of the antiprotons
rapidly attains thermal equilibrium with the positrons (Region~II) with
a time constant of $\tau \approx 10$~ms. This time corresponds to the
observed onset of high-rate production of antihydrogen. On a much
longer time scale of about~50~s, another large fraction of the injected
$\bar{p}$ cools into Region~II. The first is attributed to antiprotons
which initially overlap with the positrons and which are cooled very
efficiently in Coulomb collisions. The second cooling phase is probably
due to antiprotons which have a larger initial radius and interact only
with a thin tail of the positron plasma or possibly with residual-gas
ions surrounding it. A population of Region~III, below the positron
energy, is also observed. This is probably due to $\bar{p}$--$\bar{p}$
collisions in the lateral wells that transfer longitudinal to radial
momentum or re-ionized Rydberg $\overline{\mbox{H}}$ atoms. In the
absence of positrons, no cooling takes place and the $\bar{p}$ cloud
still has the initial kinetic energy even after 190~s.

\subsection{Recombination process}

The process that leads to the recombination of antiprotons and
positrons for the formation of antihydrogen is of the utmost
importance. It has crucial implications both for the optimization of
antihydrogen production and for the internal (atomic) states of the
produced anti-atoms. The capture of a positron by an antiproton into a
bound atomic state cannot simultaneously conserve energy and momentum;
the excess momentum has to be carried away by a third reaction partner.
The dominant processes for $\overline{\mbox{H}}$ production are
believed to be three-body recombination (TBR) and spontaneous radiative
recombination (SRR), in which this third particle is a second positron
or a photon, respectively~\cite{bib:glin1991,bib:stev1975}. These two
processes exhibit some very different properties, summarized in
Tab.~\ref{tab:recombination}.
\begin{table}
  \vspace{\tableabovecaptionadjust}%
  \caption{Comparison of some of the properties of three-body
recombination and spontaneous radiative recombination, believed to be
the dominant processes for antihydrogen production at ATHENA. The SRR
rate was calculated for $T = 15$~K, $n_{e^{+}} = 2 \times
10^{8}$/cm$^{3}$, and $N_{\bar{p}} = 10^{4}$~\cite{bib:amor2004d}.%
  \label{tab:recombination}}
  \vspace{\tablebelowcaptionadjust}
  \centering
    \begin{tabular}{lcc}
~ & Three-body recombination & Spont. radiative recombination\\
\hline%
\rule{0pt}{3ex}Temperature dep. & $\propto T^{-9/2}$ & $\propto T^{-0.63}$\\
\rule{0pt}{2.5ex}Positron density dep. & $\propto n_{e^{+}}^{2}$ & $\propto n_{e^{+}}$\\
\rule{0pt}{2.5ex}Cross-section at 1~K & $10^{-7}$~cm$^{2}$ & $10^{-16}$~cm$^{2}$\\
\rule{0pt}{2.5ex}Final internal states & $n \gg 10$ & $n < 10$\\
\rule{0pt}{2.5ex}Expected rates & unknown & $\approx40$~Hz
    \end{tabular}
\end{table}
Since high-precision measurements have to be carried out on anti-atoms
in the ground state, the most important difference between the
production mechanisms is in the final atomic states that they populate.
While SRR creates final states with low principal quantum numbers, TBR
produces Rydberg atoms with $n \gg 10$. These excited states are
expected to decay very rapidly to the ground state, but they can also
easily be re-ionized in the strong electric fields present in the
mixing region. Under real experimental conditions, an equilibrium
between de-excitation and re-ionization will probably set in.

\begin{figure}
  \centering
  \includegraphics{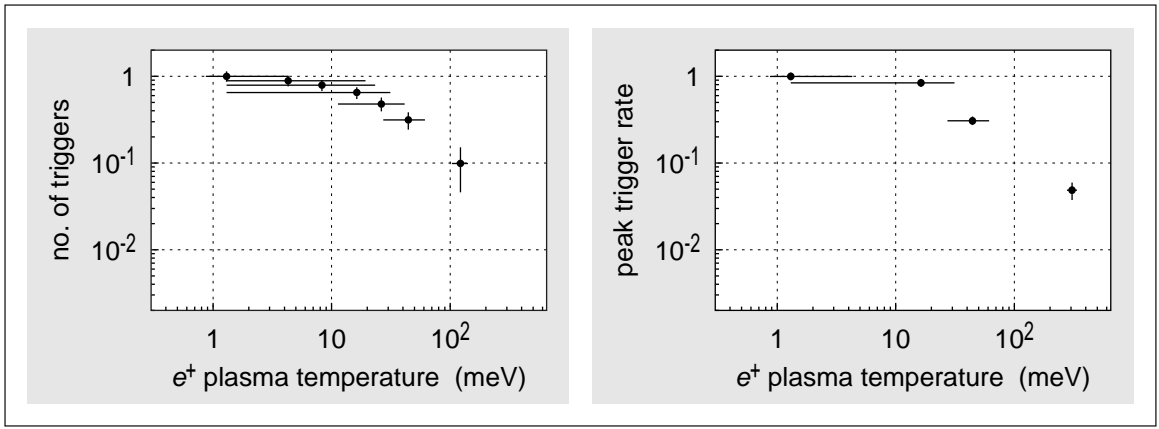}%
  \caption{Dependence of the background-corrected integrated total\
number of charged-particle triggers per mixing cycle (left) and the
peak trigger rate (left) on the positron
plasma temperature~\cite{bib:amor2004d}. Note the logarithmic scale.%
\label{fig:Temp_dependence}}
\end{figure}

As is also shown in the table, the rates of the two recombination
mechanisms follow different power laws as a function of temperature and
positron density. We have recently studied the temperature dependence
of the recombination rate in order to determine the dominant
recombination process under the experimental conditions of
ATHENA~\cite{bib:amor2004d}. We performed antiproton--positron mixing
under otherwise standard conditions, but varying the positron
temperature by RF heating of the axial plasma modes. From these data we
extracted the opening angle excess, the total number of triggers in a
complete mixing cycle (background-corrected), and the peak trigger rate
as possible proxies for the instantaneous recombination rate. The
latter two are shown in Fig.~\ref{fig:Temp_dependence}.
All of these variables show a very similar temperature dependence: The
production decreases strongly with increasing temperature (as expected
from the power laws given in Tab.~\ref{tab:recombination}), but there
is also a turnover for the lowest temperatures and the overall shape of
the graphs does not correspond to the straight line expected for a
simple power law (note the logarithmic scale). More detailed
calculations of the recombination rates, taking into account the
magnetic and electrostatic fields in the recombination region, may
eventually explain this intriguing behavior. A fit with a power law of
the form $T^{\alpha}$ to the peak trigger rate data yields $\alpha =
-0.7(2)$, in close agreement with that expected for spontaneous
radiative recombination.

\section{Conclusions and Outlook}

ATHENA can routinely produce several hundred cold antihydrogen atoms
per minute. These numbers allow detailed studies of the properties of
the produced $\overline{\mbox{H}}$, such as the process responsible for
antihydrogen recombination. The temperature dependence of the
antihydrogen production rate has been studied quantitatively. A best
fit to the data suggests spontaneous radiative recombination as the
dominant formation process, even though the observed rates are at least
an order of magnitude larger than predicted. An ongoing analysis of the
2003 data with respect to the positron density dependence should shed
more light on this important question. Notwithstanding the already
large antihydrogen production efficiency (with respect to captured
antiprotons from the AD) of about 15\%, a more complete understanding
of the recombination process and a further increase of its efficiency
are important. For this purpose, efforts are underway to enhance
$\overline{\mbox{H}}$ production by laser-stimulated recombination.
This will also constitute a first spectroscopic measurement on
antihydrogen. Furthermore, it is planned to employ a radiofrequency
excitation of the antiprotons' characteristic motions in the Penning
trap in order to center them and enhance the transfer efficiency from
the capture trap to the mixing trap and to improve the radial overlap
between the antiprotons with the positron plasma.

\section*{Acknowledgments}

This work was supported by the funding agencies INFN (Italy), CNPq
(Brazil), MEXT (Japan), SNF (Switzerland), SNF (Denmark), and EPSRC
(United Kingdom).

\end{document}